\documentstyle[preprint,aps,epsf]{revtex}

\newcommand{\be}{\begin{equation}}
\newcommand{\ee}{\end{equation}}

\begin{document}

\title{ Invalidity of Classes of Approximated Hall Effect
         Calculations }

\author{ P. Ao }

\address{ Departments of Theoretical Physics, 
Ume\aa{\ }University,
    S-901 87, Ume\aa, Sweden}


\maketitle

\noindent
pacs\#: {74.60.Ge; 67.57.Fg}


In the mixed state of a type II superconductor,
the steady limit DC resistivities are controlled by motions 
of vortices via the Josephson-Anderson relation.
The effective equation of motion of vortices has to be established
from the microscopic Bardeen-Cooper-Schrieffer theory.
In this comment, I point out a number of approximated derivations
for the effective equation of motion,
now been applied to d-wave superconductors in Ref.\cite{kv}
(see citations there for earlier references),
are invalid. The major error in those approximated derivations is the
inappropriate use of 
the relaxation time approximation in force-force correlation functions, 
or in force balance equations, or in similar variations,

The usual way of calculating DC resistivities in a condensed medium 
is simply through the inverse of the conductivity tensor. 
The AC conductivity tensor is calculated using the Nakano-Kubo formula, 
where the current-current correlation functions are calculated
with a given effective Hamiltonian.\cite{kubo}
In this scheme the relaxation time approximation can be valid\cite{kubo,zhu}.
In the superconducting state, 
however, it has been known since sixties that 
such a calculation is unattainable  in the low frequency limit, 
where the dissipation is dominated by the motion of vortices:
If all vortices are pinned down, there is no steady state solution 
for a constant applied electric field in the linear response regime.
DC resistivities have to be calculated directly, and the equation of 
motion for vortices has to be established first.
Although ingenious guesses have been made in sixties, 
with the topological methods the vortex dynamic equation has only 
been recently put on 
a solid microscopic base.\cite{at,tan,az}.

Why the microscopic derivation has taken such a long time to be finished?
The difficulty may lie in a subtlety in the direct calculation of 
DC resistivities, 
where force-force correlation functions, or force balance equations, have 
to be used, 
contrast to the current-current (or velocity-velocity)
correlation functions for the conductivities.
The random force-force correlation function gives rise directly to the
friction, or the inverse of the  relaxation time of the 
system.\cite{kubo,zhu,az}
A very well known approximation, the relaxation time approximation
 which works well in the calculation of conductivities, 
does not work in the direct calculation of DC resistivities.
Its use in force-force correlation functions or force balance equations, 
in the formulation of either
Green's function, kinetic equation, Langevin equation, or, 
path integral, 
leads to the violation of the fluctuation-dissipation theorem, 
and is therefore invalid\cite{kubo,zhu,bh,hc,fm}. 
This no-go theorem does not appear to be widely known.
 
Having discussed the invalidity of the results in Ref.\cite{kv},
let me turn to a technical error. 
Although the proposed kinetic equation, Eq.(2),
might be valid, the force balance type equation, 
Eq.(6) or (7) of Ref.\cite{kv}, for the forces from the environment 
is not. 
If the environmental degrees of freedoms for a vortex were 
the fermionic ones, the integration out the fermionic degrees of freedoms
amounts to  obtain the Feynman influence functional\cite{leggett}.
Both the transverse force and the friction on a vortex have been obtained
in this way\cite{az},
with the former independent of details,
 or in a more formal approach.\cite{tan}
There is no need of the relaxation time approximation.
If there were other dissipative mechanisms represented by the single
relaxation time in Eq.(2), 
the correct expression of obtaining forces from the environment may be found 
in a standard reference.\cite{ziman}  

To conclude, vortex dynamics results 
obtained by the relaxation time approximation in the force balance equations
of Ref.\cite{kv} are incorrect.

This work was financially supported by the Swedish NFR.

\end{document}